\documentclass[3p,times,twocolumn]{elsarticle}

\usepackage{ecrc}

\volume{00}

\firstpage{1}

\journalname{Nuclear Physics B Proceedings Supplement}

\runauth{}

\jid{nuphbp}

\jnltitlelogo{Nuclear Physics B Proceedings Supplement}

\usepackage{amssymb}

\usepackage[figuresright]{rotating}

\begin{document}

\begin{frontmatter}



\dochead{}

\title{Elastic $Z^0$ production at HERA}


\author{L. Stanco, on behalf of the ZEUS Collaboration}

\address{INFN--Padova, Via Marzolo, 8 I-35131 Padova, Italy}

\begin{abstract}
\noindent The production of $Z^{0}$ bosons in the reaction $ep\rightarrow eZ^{0}p^{\left(*\right)}$, where $p^{\left(*\right)}$ stands for a proton 
or a low-mass nucleon resonance, has been studied in $ep$ collisions at HERA using the ZEUS detector.  
The analysis is based on a data sample collected between 1996 and 2007, amounting to 496\, pb$^{-1}$ of integrated luminosity.
The $Z^{0}$  was measured in the hadronic decay mode.
The elasticity of the events was ensured by a cut on  
$\eta_{{\rm max}} < 3.0$, where $\eta_{{\rm max}}$ is the maximum pseudorapidity of energy deposits in the calorimeter 
defined with respect to the proton beam direction.
A signal was observed at the $Z^{0}$  mass.
The cross section of the reaction $ep \rightarrow eZ^{0}p^{\left(*\right)}$ was measured to be 
$\sigma \left( ep \rightarrow eZ^{0}p^{\left(*\right)} \right) = {\rm 0.13 \pm{0.06} \left( {\rm stat.} \right) \pm{0.01} \left( {\rm syst.} \right) }\, {\rm pb}$,
in agreement with the Standard Model prediction of $0.16\, {\rm pb}$.
This is the first measurement of $Z^{0}$ production in $ep$ collisions.
In this paper we report the already published ZEUS result by adding the sensitivities of the most recent similar results from CMS and ATLAS.
\end{abstract}


\end{frontmatter}


\section{Introduction}
\label{sec-int}
The production of electroweak bosons in $ep$ collisions is a good benchmark process for testing the Standard Model\,(SM).
Even though the expected numbers of events
for $W^{\pm}$ and $Z^{0}$ production are low, the measurement of the cross sections 
of these processes is important
as some extensions of the SM predict anomalous couplings and thus changes in
these cross sections.
A measurement of the cross section for $W^{\pm}$ production at HERA has been performed by H1 and ZEUS\,\cite{Wpaper} in events containing an isolated lepton and missing transverse momentum, 
giving a cross section
$\sigma \left(ep\rightarrow W^{\pm}X \right) = {{\rm 1.06} \pm {\rm 0.17} \left({\rm stat.} \oplus {\rm  syst.}\right) }\, {\rm pb},$ in good agreement with the SM prediction. 
The cross section for $Z^{0}$ production is predicted to be ${\rm 0.4}\, {\rm pb}$.

This paper reports on a measurement of the production of $Z^{0}$  bosons in $e^{\pm}p$  collisions using an integrated luminosity 
of about ${\rm 0.5}{\rm fb}^{-1}$.
The hadronic decay mode was chosen\footnote{The $Z^{0}$ decay into charged lepton pairs was studied
in a previous ZEUS publication\,\cite{phy.lett:b680:13-23}.} because of its large branching ratio.  
The excellent resolution of the ZEUS hadronic calorimeter makes this measurement possible.
The analysis was restricted to elastic and quasi-elastic $Z^{0}$  production in order to suppress QCD multi-jet background. 
The selected process is $ep \rightarrow eZ^{0}p^{\left( * \right)}$, where 
$p^{\left(*\right)}$ stands for a proton (elastic process) or a low-mass nucleon resonance (quasi-elastic process).

The corresponding and final result has been already published by the ZEUS Collaboration in~\cite{zeus-paper} which content
corresponds to this report with the addition of LHC comparison.

\begin{figure}[htbp]
\vfill
\begin{center}
\vskip-2cm\hskip-17mm
\includegraphics[angle=-90, width=1.2\hsize]{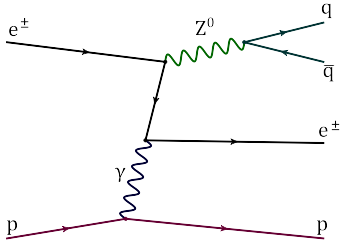}
\end{center}
\vskip-2cm
\caption{Example of a leading-order diagram of $Z^{0}$ boson production and subsequent hadronic decay (into quark $q$ and antiquark $\bar{q}$) 
in  $ep\rightarrow eZ^{0}p$. }
\label{diagram}
\vfill
\end{figure}

Fig.~\ref{diagram} shows a leading-order (LO) diagram of $Z^{0}$ production
with subsequent hadronic decay.
In such events, there are at least two hadronic jets with high transverse energies, and no hadronic energy deposits around the 
forward\footnote{The ZEUS coordinate system 
is a right-handed Cartesian system, with the $Z$ axis pointing in the proton beam direction, referred to as the forward direction, 
and the $X$ axis pointing towards the centre of HERA.
The coordinate origin is at the nominal interaction point. The pseudorapidity is defined as 
$\eta = -\ln\left( \tan{\frac{\theta}{2}} \right)$, where the polar angle, $\theta$, 
is measured with respect to the proton beam direction.} 
direction, in contrast to what would be expected in inelastic collisions.

\section{Experimental set-up}
\label{sec-exp}
HERA was the world's only high-energy $ep$ collider, with an electron\footnote{The term ``electron'' also refers to positrons if not stated otherwise.} 
beam of ${\rm 27.6\,}{\rm GeV}$ and a proton beam of 
${\rm 920\,}{\rm GeV}$\,(${\rm 820\,}{\rm GeV}$ until 1997).
For this analysis, $e^{\pm}p$ collision data collected with the ZEUS detector between 1996 and 2007, 
amounting to ${\rm 496}\, {\rm pb}^{-1}$ of integrated luminosity, have been used.
They consist of ${\rm 289}\, {\rm pb}^{-1}$ of $e^{+}p$  data and ${\rm 207}\, {\rm pb}^{-1}$ of $e^{-}p$ data.

After 2003, HERA was operated with a polarised lepton beam.
When combining the data taken with negative and positive polarisations,
the average polarisation is less than 1\% and its effect was neglected in this analysis.

A detailed description of the ZEUS detector can be found 
elsewhere~[3 in ref.~1]. 
A brief outline of the calorimeter
that is most relevant for this analysis follows.


The high-resolution uranium--scintillator calorimeter (CAL)~\cite{CAL}
consisted of three parts: the forward (FCAL), the barrel (BCAL) and
the rear (RCAL) calorimeters. Each part was subdivided transversely
into towers and longitudinally into one electromagnetic section (EMC)
and either one (in RCAL) or two (in BCAL and FCAL) hadronic sections
(HAC). The smallest subdivision of the calorimeter was called a cell.
The CAL energy resolutions, as measured under test-beam conditions,
were $\sigma(E)/E=0.18/\sqrt{E}$ for electrons and
$\sigma(E)/E=0.35/\sqrt{E}$ for hadrons, with $E$ in GeV.



The luminosity was measured using the Bethe--Heitler reaction
$ep\,\rightarrow\, e\gamma p$. The fractional systematic
uncertainty on the measured luminosity was 2\%.


\section{Monte Carlo simulations}\label{mcsimulation}
Monte Carlo (MC) simulations were made to simulate the $Z^{0}$ production process.
They were used to correct for instrumental effects and selection acceptance 
and to provide a template for the shape of the invariant-mass distribution of the $Z^{0}$  signal.
The EPVEC program~\cite{np:b375:3} was used to generate the signal events at the parton level. 
The following $Z^{0}$ production processes are considered in EPVEC:
\begin{itemize}
\itemsep-0.15cm
\item elastic scattering, $ep \rightarrow eZ^{0}p$, where the proton stays intact;
\item quasi-elastic scattering,  $ep \rightarrow eZ^{0}p^{*}$, where the proton is transformed into a nucleon resonance $p^{*}$;
\item deep inelastic scattering\,(DIS), $\gamma^{*}p \rightarrow Z^{0}X$, in the region $Q^{2} > \, {\rm 4}\,{\rm GeV}^2$, 
where $Q^{2}$ is the virtuality of the photon exchanged between the electron and proton;
\item resolved photoproduction, $\gamma p \rightarrow \left( q\bar{q} \rightarrow Z^{0} \right)X$, where one of the quarks is 
a constituent of the resolved photon and the other quark is a constituent of the proton.
\end{itemize}
In EPVEC the first two processes 
are calculated using form factors and structure functions fitted directly to experimental data. 
Note that, even if the virtuality of the exchanged photon is small, the scattered electron could receive a large momentum transfer when the 
$Z^{0}$ is radiated from the lepton line.
In the last two processes, the proton breaks up.
The DIS process is calculated in the quark--parton model using a full set of leading-order Feynman diagrams.
Resolved photoproduction is parametrised using a photon structure function and is carefully matched to the DIS region.
The cross section of $Z^{0}$ production is calculated to be ${\rm 0.16}\, {\rm pb}$ for elastic and quasi-elastic processes and ${\rm 0.24}\, {\rm pb}$ for DIS and resolved photoproduction.
The difference between $e^{+}p$ and $e^{-}p$ cross sections is negligible for this analysis ($<$1\% for the DIS process).
A correction, based on the MC cross section,  was made to account for the part of data taken at the centre-of-mass energy $\sqrt{s}={\rm 300\,}{\rm GeV}$, 
so that the result is quoted at $\sqrt{s}={\rm 318\,}{\rm GeV}$.

After the parton-level generation by EPVEC, PYTHIA 5.6~\cite{manual:cern-th-6488/92} was used to simulate initial- and final-state parton showers with the fragmentation into hadrons using the Lund string model~\cite{prep:97:31} as implemented in JETSET 7.3~\cite{manual:cern-th-6488/92}.
The generated MC events were passed through the ZEUS detector
and trigger simulation programs based on GEANT 3.13~\cite{tech:cern-dd-ee-84-1}.
They were reconstructed and analysed by the same programs as the data.

A reliable prediction of background events with the signal topology, which are predominantly due to the diffractive photoproduction 
of jets of high transverse momentum, 
is currently not available.
Therefore, the background shape of the invariant-mass distribution was estimated with a data-driven method.
The normalisation was determined by a fit to the data.

\section{Event reconstruction and selection}
\label{obj_recon}

The events used in this analysis were selected online by the ZEUS three-level trigger system~[13 in ref.~1], 
using a combination of several trigger chains which 
were mainly based on requirements of large transverse energy deposited in the calorimeter.
In the offline selection, further criteria were imposed in order to separate the signal from the background.

The events are characterised by the presence of at least two jets of high transverse energy
and, for a fraction of events, by the presence of a reconstructed scattered electron.
In order to select events with a $Z^{0}$ decaying hadronically, jets were reconstructed in the hadronic final state using the $k_{T}$ cluster 
algorithm~\cite{np:b406:187} in the longitudinally invariant inclusive mode~\cite{pr:d48:3160}.
The algorithm was applied to the energy clusters in the CAL after excluding those associated with the scattered-electron 
candidate~\cite{epj:c11:427,thesis:briskin:1998,Wai:1995cx}. 
Energy corrections~[19--21 in ref.~1] 
were applied to the jets in order to compensate for energy 
losses in the inactive material in front of the CAL.

In this analysis, only jets with 
$E_{T} > \, {\rm 4\,}{\rm GeV}$ and $\left| \eta \right| < 2.0$ were used. 
Here  $E_{T}$ is the jet transverse energy and $\eta$ its pseudorapidity.
The hadronic $Z^{0}$ decay sample was selected by the following requirements on the reconstructed jets:
\begin{itemize}
\itemsep-0.15cm
\item at least two jets in the event had to satisfy $E_{T} > \, {\rm 25\,}{\rm GeV}$;
\item $\left| \Delta \phi_{j} \right| > \, {\rm 2}\, {\rm rad}$, where $\Delta \phi_{j}$ is the azimuthal difference between the first and second 
highest-$E_{T}$ jet, as the two leading jets from the $Z^{0}$ boson decays are expected to be nearly back-to-back in the  $X$--$Y$ plane.
\end{itemize}

Electrons were reconstructed using an algorithm that combined information from clusters of energy deposits in the CAL and from tracks~\cite{epj:c11:427}. 
To be defined as well-reconstructed electrons, the candidates were required to satisfy the following selection:
\begin{itemize}
\itemsep-0.15cm
\item $ E^{'}_e > \, {\rm 5\,}{\rm GeV}$ and $ E_{{\rm in}} < \, {\rm 3\,}{\rm GeV}$, where $ E^{'}_{e}$ is the scattered electron energy and
$ E_{{\rm in}}$ is the total energy in all CAL cells not associated with the cluster of the electron but lying within a cone 
in $\eta$ and $\phi$ of radius $R = \sqrt{ \Delta \eta^{2} + \Delta \phi^{2} } = 0.8$, centred on the cluster;
\item If the electron was in the acceptance region of the tracking system,
a matched track was required with momentum $p_{{\rm track}} > \, {\rm3\,}{\rm GeV}$.
After extrapolating the track to the CAL surface, its distance of
closest approach (DCA) to the electron cluster had to be within 10 cm. 
\end{itemize}

The following cuts were applied to suppress low-$Q^{2}$ neutral-current and direct-photoproduction backgrounds:
\begin{itemize}
\itemsep-0.15cm
\item $E_{{\rm RCAL}} < \, {\rm 2\,}{\rm GeV}$, where $E_{{\rm RCAL}}$ is the total energy deposit in RCAL;
\item $50<E-p_{Z}<\, {\rm 64\,}{\rm GeV}$,
 where $E-p_{Z} = \sum_{i} E_{i} \left( 1-\cos\theta_{i} \right)$;
$E_{i}$ is the energy of the $i$-th CAL cell, $\theta_{i}$ is its polar angle and the sum runs over all 
cells\footnote{For fully contained events, or events in which the particles escape only in the forward beam pipe, the $E-p_{Z}$ value peaks around twice
the electron beam energy, 55 GeV.};
\item $\theta_{e} < 80^{\circ}$ for well reconstructed electrons, where $\theta_{e}$ is the polar angle of the scattered electron,
motivated by the fact that, due to the large mass of the produced system, the electron is backscattered to the forward calorimeter or forward beam pipe;
\item the event was rejected if more than one electron candidate was found;
\item jets were regarded as a misidentified electron or photon and were discarded from the list of jets if the direction of the jet 
candidate was matched within $R < 1.0$ with that of an electron candidate identified by looser criteria\footnote{
Candidates were selected by less stringent requirements and clusters with no tracks were also accepted to find photons and electrons.}
than those described above.
This cut causes a loss of acceptance of about 3\%.
\end{itemize}

To remove cosmic and beam--gas backgrounds, events fulfilling any of the conditions listed below were rejected:
\begin{itemize}
\itemsep-0.15cm
\item $|Z_{{\rm vtx}}| > \, {\rm 50}\, {\rm cm} $, where $Z_{{\rm vtx}}$ is the $Z$ position of the primary vertex reconstructed from CTD+MVD tracks;
\item $175^{\circ} < \left( \theta_{{\rm jet}1} + \theta_{{\rm jet}2} \right) < 185^{\circ}$ and $\Delta \phi_{j} > 175^{\circ}$ simultaneously,
 where $\theta_{{\rm jet}1}$ and $\theta_{{\rm jet}2}$ are the polar angles of the first and second highest-$E_{T}$  jet, respectively, 
and $\Delta \phi_{j}$ is the azimuthal difference between them;
\item $\left| t_{{u}} - t_{d} \right| > \, {\rm 6.0}\, {\rm ns}$, where $\left| t_{u} - t_{d} \right|$ is  
the timing difference between the upper and the lower halves of the BCAL;
\item $p\hspace{-.47em}/_{T} > \, {\rm 25\,}{\rm GeV}$, where $p\hspace{-.47em}/_{T}$ is the missing transverse momentum calculated from the energy clusters in the CAL;
\item $N_{{\rm trk}}^{{\rm vtx}} < 0.25 \left( N_{{\rm trk}}^{{\rm all}} - 20 \right)$, where $N_{{\rm trk}}^{{\rm vtx}}$ is
the number of tracks associated with the primary vertex and $N_{{\rm trk}}^{{\rm all}}$ is the total number of tracks~\cite{pl:b539:197}.
\end{itemize}

The number of events passing the above selection was 5257. 
Finally, to select the elastic and quasi-elastic processes preferentially, a cut on $\eta_{{\rm max}}$ was introduced,
\begin{itemize}
\itemsep-0.15cm
\item $\eta_{{\rm max}} < 3.0$.
\end{itemize}
The quantity $\eta_{{\rm max}}$ was defined as the pseudorapidity of the energy deposit in the calorimeter closest to the 
proton beam direction with energy greater than ${\rm 400\,}{\rm MeV}$ as determined by calorimeter cells.
This cut also rejected signal events which have energy deposits from the scattered
electron in the calorimeter around the forward beam pipe, causing an acceptance loss of about 30\%.

After all selection cuts, 54 events remained.
The total selection efficiency was estimated by the MC simulation to be $22\%$ for elastic and quasi-elastic processes and less than $1\%$ for DIS and resolved photoproduction events.
The number of expected signal events in the final sample, as predicted by EPVEC,  is 18.3.
The contribution from elastic and quasi-elastic processes amounts to 17.9 events.

\section{Background-shape study}
\label{bgshape}
\begin{figure}[htbp]
\vfill
\begin{center}
\includegraphics[width=\hsize]{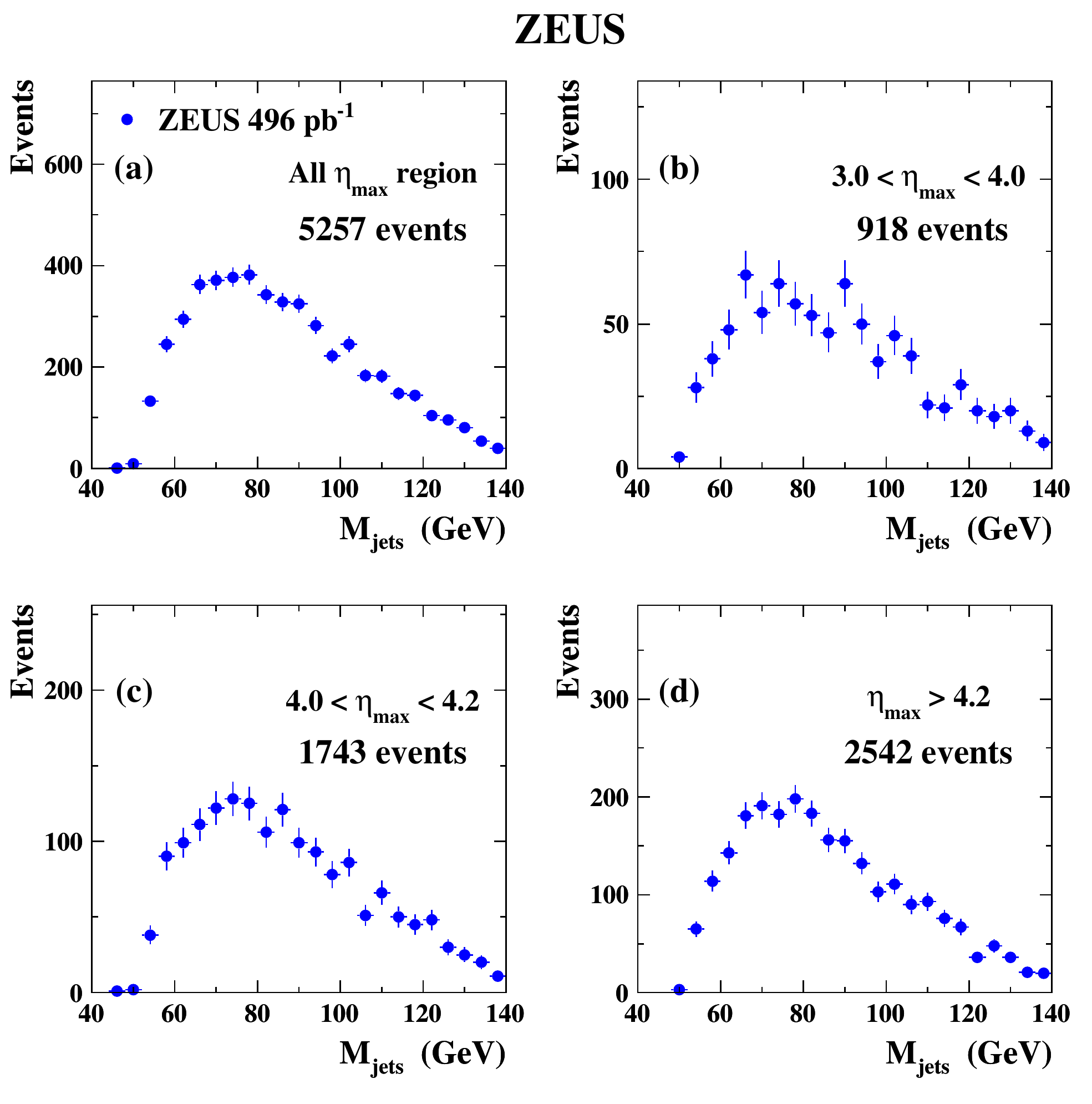}
\end{center}
\caption{The $M_{{\rm jets}}$ distribution of the data (a) after all selection criteria, except for the $\eta_{{\rm max}}$ cut,  
(b-d) in several $\eta_{{\rm max}}$ slices.}
\label{dataall}
\vfill
\end{figure}

\begin{figure}[htbp]
\vfill
  \begin{center}
\includegraphics[width=1.0\hsize]{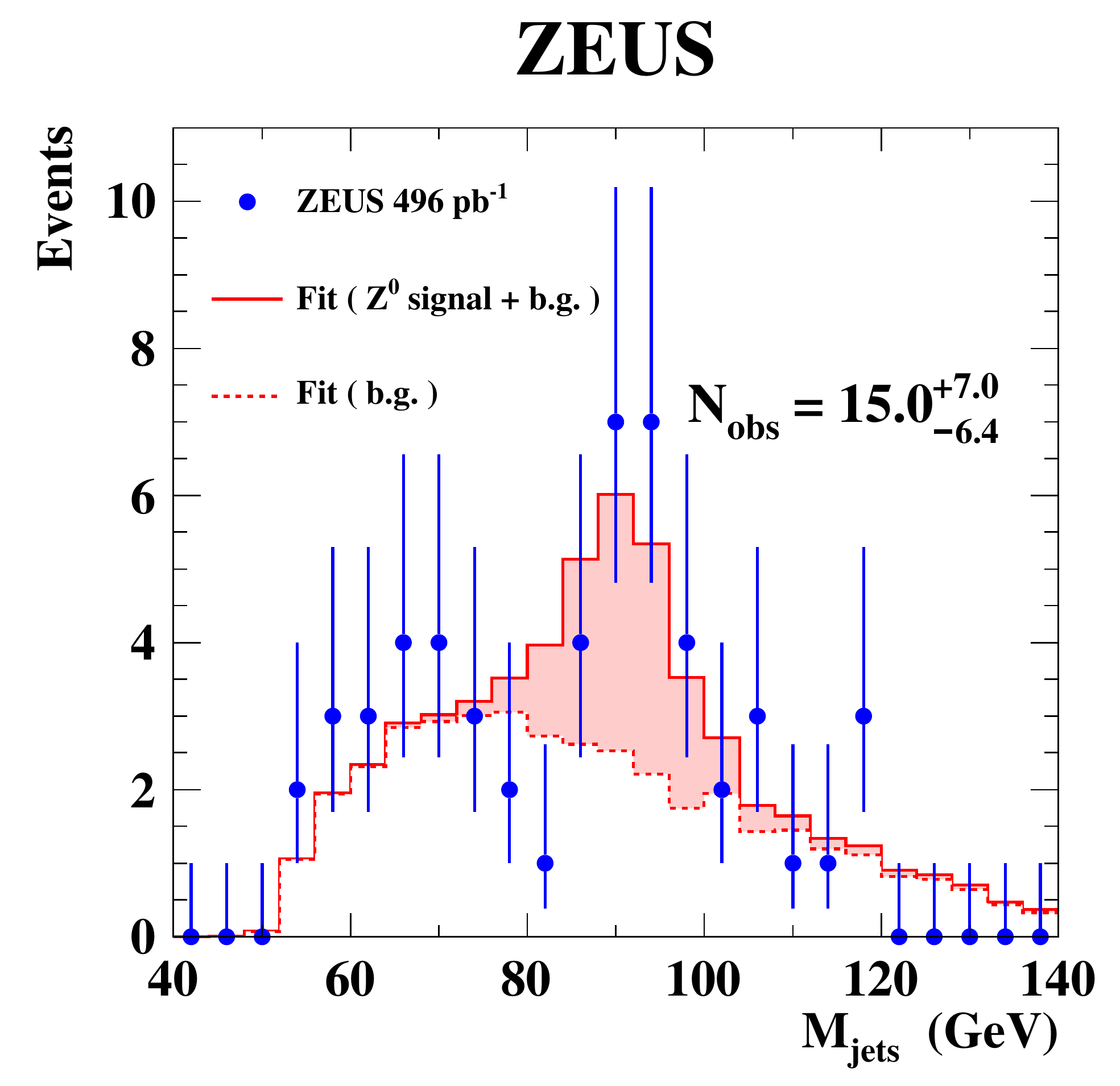}
    \caption{The $M_{{\rm jets}}$ distribution and the fit result. The data are shown as points, and the fitting result of signal+background\,(background component) 
is shown as solid\,(dashed) line. 
The signal contribution is also indicated by the shaded area and amounts
to a total number of $N_{obs}$ events.
The error bars represent the approximate Poissonian 68\% CL intervals, calculated as $\pm \sqrt{n+0.25} + 0.5$ for a given entry $n$.}
    \label{mass_final}
  \end{center}
\vfill
\end{figure}

Fig.~\ref{dataall}(a) shows the distribution of the invariant mass, 
$M_{{\rm jets}}$, 
after all the selection criteria except for the requirement $\eta_{{\rm max}} < 3.0$.
The variable $M_{{\rm jets}}$ was calculated using all jets passing the selection criteria described in Section\,\ref{obj_recon}.
Figs.~\ref{dataall}(b)-(d) show $M_{{\rm jets}}$ for various $\eta_{{\rm max}}$ slices 
in the inelastic region\,($\eta_{{\rm max}} > 3.0$) for the same selection.
No significant dependence on $\eta_{{\rm max}}$ of the $M_{{\rm jets}}$ distribution beyond that expected from statistical 
fluctuations was observed in the inelastic region.
In addition, the shape of the $M_{{\rm jets}}$ distribution outside the $Z^{0}$ mass window in the region $\eta_{{\rm max}} < 3.0$ 
was found to be consistent with that in the inelastic region\,(Fig. \ref{mass_final}).
Therefore, the $M_{{\rm jets}}$ distribution in the inelastic region was adopted as a background template in a fit to the data 
in the elastic region as described in the following section.

\section{Cross-section extraction}
A fit to the sum of the signal and a background template for the $M_{{\rm jets}}$ distribution was used for the cross-section extraction.
The template $N_{{\rm ref},i}$ is defined according to:
\begin{equation}
\hskip-0.75cm
N_{{\rm ref},i} = aN_{{\rm sg},i}^{{\rm MC}}\left( \epsilon \right) + bN_{{\rm bg}, i}^{{\rm data}},
\label{eq1}
\end{equation}
where $i$ is the bin number of the $M_{{\rm jets}}$ distribution. The parameter $\epsilon$ accounts for a possible energy shift, i.e. 
$M_{{\rm jets}}=\left( 1+\epsilon \right)M_{{\rm jets}}^{{\rm MC}}$, where $M_{{\rm jets}}^{{\rm MC}}$ is the invariant-mass distribution 
of the signal $Z^{0}$ MC.
The quantity $N_{{\rm sg},i}^{{\rm MC}}$ is a signal template estimated from the $Z^{0}$ MC distribution after all cuts, 
normalised to data luminosity. 
The quantity  $N_{{\rm bg},i}^{{\rm data}}$ is a background template determined from the data outside the selected region.
The parameters $a$ and $b$ are the normalisation factors for the signal and background, respectively.
The likelihood of the fit, $\mathcal{L}$, is defined as follows:
\begin{equation}
\hskip-0.75cm
  \mathcal{L} =  \mathcal{L}_{1} \left( N_{{\rm obs}}, N_{{\rm ref}} \right)  \times \mathcal{L}_{2}\left( \epsilon, \sigma_{\epsilon} \right), \\
\label{eq2}
\end{equation}
with
{\small
\[
\hskip-0.75cm
\mathcal{L}_{1} = \prod_{i} \frac{ {\rm exp} \left( -N_{{\rm ref}, i} \right)\left( N_{{\rm ref}, i} \right)^{N_{{\rm obs}, i}} }{ N_{{\rm obs}, i}! } {\rm \ \ \ ~and~ \ \ \ } \mathcal{L}_{2} = {\rm exp}\left( -\frac{\epsilon^{2}}{2\sigma_{\epsilon}^{2}} \right).  \nonumber
\]
}
Here $\mathcal{L}_{1}\left( N_{{\rm obs}}, N_{{\rm ref}} \right)$ is the product of Poisson probabilities to observe $N_{{\rm obs}, i}$ events
for the bin $i$ when $N_{{\rm ref}, i}$ is expected.
The term  $\mathcal{L}_{2}\left( \epsilon, \sigma_{\epsilon} \right)$ represents the Gaussian probability density for a shift $\epsilon$ of the jet energy scale from the nominal scale,
which has a known systematic uncertainty of $\sigma_{\epsilon}=3\%$.
From the likelihood, a chi-squared function  is defined as
\begin{equation}
\hskip-0.75cm
\tilde{ \chi }^{2} = -2\ln \frac{\mathcal{L}_{1}\left( N_{{\rm obs}}, N_{{\rm ref}} \right)}{\mathcal{L}_{1}\left( N_{{\rm obs}}, N_{{\rm obs}} \right)} - 2\ln\mathcal{L}_{2} = 2\sum f_{i}+ \left( \frac{\epsilon}{\sigma_{\epsilon}} \right)^{2}, \\
\end{equation}
with
{\small\[
\hskip-0.75cm
f_{i} =\left\{ \begin{array}{l}
				N_{{\rm ref},i} - N_{{\rm obs},i} + N_{{\rm obs},i}\ln\left( N_{{\rm obs},i}/N_{{\rm ref},i} \right)
				\hspace{5.85mm}\left( {\rm if}\,N_{{\rm obs},i}>0 \right) \\
				N_{{\rm ref},i}  
				\hspace{44.5mm}\left( {\rm if}\,N_{{\rm obs},i}=0 \right).
				\end{array} 
				\right.\nonumber
\]}
The best combination of ($a$,$b$,$\epsilon$) is found by minimising $\tilde{ \chi }^{2}$.
The value of $a$ after this optimisation gives the ratio between the observed and expected cross section, i.e. 
$\sigma_{{\rm obs}}=a \sigma_{{\rm SM}}$.
The maximum and minimum values of $a$ in the interval $\Delta \tilde{ \chi }^{2} < 1$ define the range of statistical uncertainty.

\section{Systematic uncertainties}
Several sources of systematic uncertainties were considered and their impact on the measurement estimated.
\begin{itemize}
\itemsep-0.15cm
\item An uncertainty of 3\% was assigned to the energy scale of the jets and the effect on the acceptance correction was estimated using the signal MC.
The uncertainty on the $Z^{0}$ cross-section measurement was estimated to be $+2.1\%$ and $-1.7\%$.
\item The uncertainty associated with the elastic and quasi-elastic selection was considered.
In a control sample of diffractive DIS candidate events, the $\eta_{{\rm max}}$ distribution of the MC agreed with the data to within 
a shift of $\eta_{{\rm max}}$ of 0.2 units~\cite{thesis:sola:2012}.
Thus, the $\eta_{{\rm max}}$ threshold was changed in the signal MC by $\pm 0.2$, and variations of the acceptance were calculated accordingly.
The uncertainty on the cross-section measurement was $+6.4\%$ and $-5.4\%$.
\item  The background shape uncertainty was estimated by using different slices of $\eta_{{\rm max}}$ in the fit.
The background shape was obtained using only the regions of $4.0 < \eta_{{\rm max}} < 4.2$ or $4.2 < \eta_{{\rm max}}$.
The region of  $3.0 < \eta_{{\rm max}} < 4.0$ was not used since a small number of signal events is expected in this $\eta_{{\rm max}}$ 
region\footnote{The ratio of the expected number of signal MC events to the observed data in this region was estimated to be 2.6\% 
for $80 < M_{{\rm jets}} < 100$\,GeV, while in the other slices it was less than 0.4\%.}.
The resulting uncertainty in the cross-section measurement was $\pm1.5$\%.
\item The uncertainty associated with the luminosity measurement was estimated to be 2\%, as described in Section \ref{sec-exp}.
\item The $Z^{0}$ mass distribution from the MC used as a signal template has a Gaussian core width of 6~GeV. 
A possible systematic uncertainty coming from the width of the MC signal peak was studied. 
The mass fit was repeated after smearing the $Z^{0}$ mass distribution in the MC by a Gaussian function with different widths. 
The measured cross section did not change significantly after smearing the distribution up to the point where
the fit  $\tilde{ \chi }^{2}$ changed by 1. No systematic uncertainty from this source was assigned.
\end{itemize}
The total systematic uncertainty was calculated by summing the individual uncertainties in quadrature and amounts to $+7.2\%$ and $-6.2\%$.

\section{Results and conclusions}
Fig.~\ref{mass_final} shows the invariant-mass distribution of the selected events.
It also shows the fit result  
for the signal plus background and the background separately.
As described in the previous sections, the background is mainly from diffractive multi-jet production, 
and the template of its invariant-mass distribution is determined from the data. 
The fit yielded a result for the parameter $a$ from Eq.~\ref{eq1} of $a = 0.82^{+0.38}_{-0.35}$.
That translates into a number of observed $Z^{0}$ events of $15.0 ^{+7.0}_{-6.4}$~(stat.), which corresponds
to a signal with a $2.3 \sigma$ statistical significance.
The fit yielded a value for $\epsilon$, the potential energy shift with respect to the signal MC,
of $0.028^{+0.021}_{-0.020}$, which is compatible with zero.
The correlation between the parameters $a$ and $\epsilon$ is rather weak; when fixing the value of $\epsilon$ 
to zero the minimum $\tilde{ \chi }^{2}$ is observed at $a = 0.65$. 
The quality was evaluated according to Eq. 3; the value of $\tilde{ \chi }^{2} / {\rm ndf}= 17.6 / 22$, 
where ndf is the number of degrees of freedom, indicates a good fit.
The cross section for the elastic and quasi-elastic production of $Z^{0}$ bosons, 
$ep \rightarrow eZ^{0}p^{\left( * \right)}$,  at $\sqrt{s} = \, {\rm 318\,}{\rm GeV}$, was calculated to be
\[
\hskip-0.75cm
\sigma(ep \rightarrow eZ^{0}p^{\left( * \right)}) = \, {\rm 0.13 \pm {0.06} \left( {\rm stat.} \right) \pm{0.01} \left( {\rm syst.} \right) }\, {\rm pb}.
\]
\noindent This result is consistent with the SM cross section calculated with EPVEC of ${\rm 0.16}\, {\rm pb}$.
This represents the first observation of $Z^{0}$ production in $ep$ collisions.

\section{Hadronic $Z^0$ mass resolutions at LHC}\label{lhc-res}
\begin{figure}[htbp]
\vfill
\begin{center}
\includegraphics[scale=0.6]{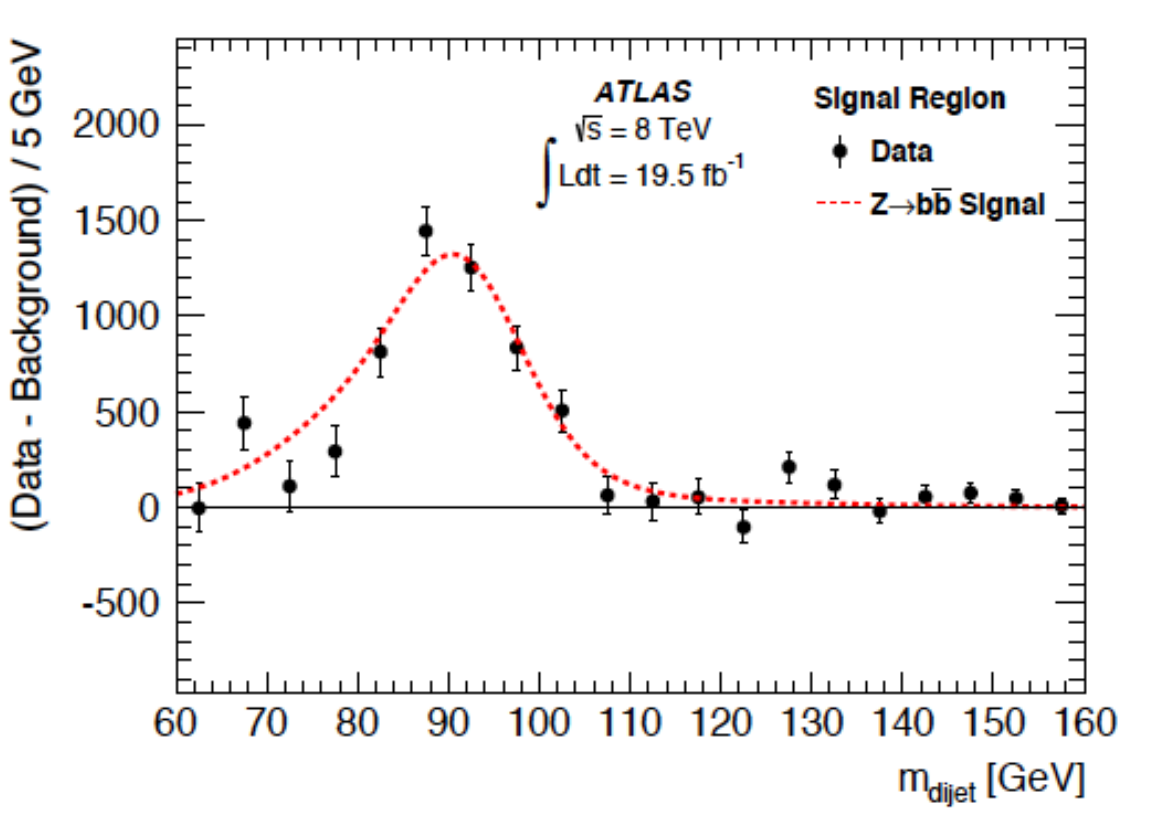}
\end{center}
\vskip-0.5cm
\caption{The ATLAS di--jet mass distribution of the selected events for $Z^0\rightarrow b\overline{b}$~\cite{ref:atlas} search.}
\label{fig:atlas}
\vfill
\end{figure}

\begin{figure}[htbp]
\vfill
\begin{center}
\includegraphics[scale=0.6]{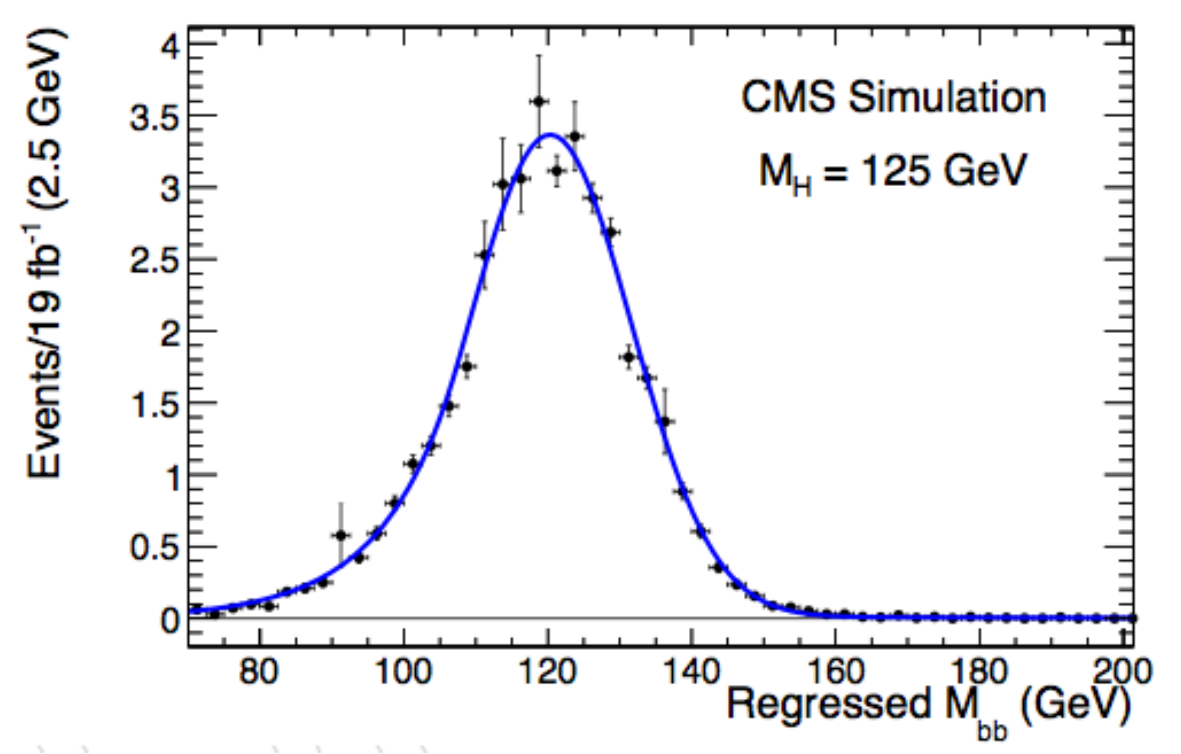}
\end{center}
\vskip-0.5cm
\caption{The CMS di--jet mass template for the event selection of $H\rightarrow b\overline{b}$~\cite{ref:cms} 
(courtesy of the CMS Collaboration).}
\label{fig:cms}
\vfill
\end{figure}

The excellent resolution obtained for the jet mass at the $Z^0$ scale is the best achieved so far in similar apparata. That is due
to the excellent performances of the ZEUS hadronic calorimeter. By comparison we report the recent achievements from ATLAS and CMS
experiments. For example, the result by ATLAS for the mass resolution obtained in 
$Z^0\rightarrow b\overline{b}$~\cite{ref:atlas} is shown in Fig.~\ref{fig:atlas}. 
Similarly, in the CMS experiment the expected resolution on the di--jet mass scale is
shown in Fig.~\ref{fig:cms} for the $H\rightarrow b\overline{b}$ decay~\cite{ref:cms}. In either case the resolution comes about a factor two worse
than that obtained by the ZEUS calorimeter.

\section*{Acknowledgements}
\label{sec-ack}

We appreciate the contributions to the construction and maintenance of
the ZEUS detector of many people who are not listed as authors. The
HERA machine group and the DESY computing staff are especially
acknowledged for their success in providing excellent operation of the
collider and the data-analysis environment. We thank the DESY
directorate for their strong support and encouragement.\\
We also thank the organizers of ICHEP2014 for the kind invitation to report
about this final result from the ZEUS Collaboration.





\nocite{*}
\bibliographystyle{elsarticle-num}
\bibliography{martin}



\end{document}